\documentstyle[aps,prc]{revtex}
\draft

\newcommand{\ba}{\begin{eqnarray}}
\newcommand{\ea}{\end{eqnarray}}
\input{psfig}
\def\ii{\'{\i}}

\begin{document}
\pagestyle{plain}

\title{Electromagnetic form factors of baryons 
in an algebraic approach}

\author{R. Bijker}
\address{Instituto de Ciencias Nucleares, 
        Universidad Nacional Aut\'onoma de M\'exico, 
        A.P. 70-543, 04510 M\'exico D.F., M\'exico}
\author{A. Leviatan}
\address{Racah Institute of Physics, 
        The Hebrew University, Jerusalem 91904, Israel}

\maketitle

\begin{abstract}
We present a simultaneous analysis of elastic and transition 
form factors of the nucleon. The calculations are performed in 
the framework of an algebraic model of baryons. Effects of 
meson cloud couplings are considered.
\\
\mbox{}

\mbox{}

Se presenta un analisis de los factores de forma electromagn\'eticos 
del nucle\'on en el contexto de un modelo algebraico. Se considere 
los efectos de los acoplamientos con la nube mes\'onica. \\
\mbox{}

Descriptores: Espectroscop\ii a bari\'onica; factores de forma 
electromagn\'eticos; m\'etodos algebraicos; dominancia de 
mesones vectoriales.

\end{abstract}


\section{Introduction}

The electromagnetic form factors of the nucleon and its excitations 
(baryon resonances) provide a powerful tool to investigate the 
structure of the nucleon \cite{Baryons}. These form factors can be 
measured in electroproduction as a function of the four-momentum 
squared $q^2=-Q^2$ of the virtual photon. 

The elastic form factors of the nucleon have mostly been 
determined with the Rosenbluth separation technique. However, for 
values of $Q^2$ above a few (GeV/c)$^2$ it becomes increasingly 
difficult to extract the electric form factor $G_E$ 
because the measured cross section is dominated by 
the magnetic form factor $G_M$. For neutron form factors 
there is the additional problem of the lack of a free neutron target, 
and the (small) relative size of the electric 
versus the magnetic form factors for small values of $Q^2$. 
With the advent of high duty cycle electron facilities such as 
ELSA, MAMI, and Jefferson Laboratory far more accurate determinations of 
the form factors become feasible (ratio method, polarization variables), 
especially for the neutron form factors 
(see {\em e.g.} \cite{gendat,gmndat2}). 
Recent experiments on electroproduction and eta-photoproduction 
have yielded valuable new information on transition form factors, 
in particular on the helicity amplitudes of the N(1520)$D_{13}$ and 
N(1535)$S_{11}$ resonances \cite{Nimai1,Nimai2,Armstrong}.

The challenge for theoretical calculations is to provide a simultaneous 
description of both the four elastic form factors and the transition 
form factors, even if it is only in a phenomenological approach. 
In this contribution we present such an analysis in the framework of a 
recently introduced algebraic model of the nucleon \cite{BIL}. 

\section{Algebraic model}

The algebraic approach provides a unified treatment of various 
constituent quark models \cite{BIL}, such as harmonic oscillator 
quark models and collective models. In this paper we employ 
a collective model of the nucleon in which baryon resonances are 
interpreted as vibrational and rotational excitations of an 
oblate top. There are two fundamental vibrations: a breathing 
mode and a two-dimensional vibrational mode, which are 
associated with the $N(1440)P_{11}$ Roper resonance and the 
$N(1710)P_{11}$ resonance, respectively. The negative parity 
resonances of the second resonance region are interpreted as 
rotational excitations. Since each vibrational mode has its own 
characteristic frequency, there is no problem with the mass of the 
Roper resonance relative to that of the negative parity resonances. 

The electromagnetic form factors are obtained by folding with 
a distribution of the charge and magnetization over the entire 
volume of the baryon \cite{BIL}. 
All calculations are carried out in the Breit frame. 

\section{Elastic form factors}

In \cite{emff} we studied the elastic electromagnetic form factors 
of the nucleon. These calculations include anomalous magnetic moments 
for the proton and the neutron, as well as a flavor dependent 
distribution functions of the charge and magnetization. 
Supposedly, the anomalous magnetic moments and the flavor dependence 
arise as effective parameters, since the coupling to the meson cloud 
surrounding the nucleon was not included explicitly. 
According to \cite{emff} the electric and magnetic form factors 
of the nucleon, when folded with a distribution of the charge 
and magnetization, can be expressed in terms of a common intrinsic 
dipole form factor
\ba
G_E^p(Q^2) &=& g(Q^2) ~,
\nonumber\\
G_E^n(Q^2) &=& 0 ~,
\nonumber\\
G_M^p(Q^2) &=& g(Q^2) ~,
\nonumber\\
G_M^n(Q^2) &=& -2g(Q^2)/3 ~,
\label{Sachs}
\ea
with
\ba
g(Q^2) &=& \frac{1}{(1+\gamma Q^2)^2} ~. 
\ea
Note that the form factors of Eq.~(\ref{Sachs}) do not contain 
anomalous magnetic moments nor involve flavor dependent distribution 
functions. In order to study the coupling to the meson cloud we express 
the electric and magnetic form factors in terms of their isoscalar ($S$) 
and isovector ($V$) components
\ba
G_{E}^S(Q^2) &=& G_{E}^p(Q^2) + G_{E}^n(Q^2) \;=\; g(Q^2) ~,
\nonumber\\
G_{E}^V(Q^2) &=& G_{E}^p(Q^2) - G_{E}^n(Q^2) \;=\; g(Q^2) ~,
\nonumber\\
G_{M}^S(Q^2) &=& G_{M}^p(Q^2) + G_{M}^n(Q^2) \;=\; g(Q^2)/3 ~,
\nonumber\\
G_{M}^V(Q^2) &=& G_{M}^p(Q^2) - G_{M}^n(Q^2) \;=\; 5g(Q^2)/3 ~. 
\ea

\subsection{Meson cloud couplings}

The effects of the meson cloud surrounding the nucleon are taken into 
account by including the coupling to the isoscalar vector mesons 
$\omega$ and $\phi$ and the isovector vector meson $\rho$. 
These contributions are studied phenomenologically by parametrizing 
the isoscalar and isovector components of the form factors as 
\ba
G_E^S(Q^2) &=& g(Q^2) \left[ \alpha^S 
+ \alpha_{\omega} \, \frac{m_{\omega}^2}{m_{\omega}^2+Q^2} 
+ \alpha_{\phi} \, \frac{m_{\phi}^2}{m_{\phi}^2+Q^2} \right] ~,
\nonumber\\
G_E^V(Q^2) &=& g(Q^2) \left[ \alpha^V  
+ \alpha_{\rho} \, \frac{m_{\rho}^2}{m_{\rho}^2+Q^2} \right] ~,
\nonumber\\
G_M^S(Q^2) &=& \frac{1}{3} g(Q^2) \left[ \beta^S 
+ \beta_{\omega} \, \frac{m_{\omega}^2}{m_{\omega}^2+Q^2} 
+ \beta_{\phi} \, \frac{m_{\phi}^2}{m_{\phi}^2+Q^2} \right] ~,
\nonumber\\
G_M^V(Q^2) &=& \frac{5}{3} g(Q^2) \left[ \beta^V  
+ \beta_{\rho} \, \frac{m_{\rho}^2}{m_{\rho}^2+Q^2} \right] ~.
\label{SV}
\ea
The large width of the $\rho$ meson ($\Gamma_{\rho}=151$ MeV) is taken 
into account by making the replacement \cite{IJL} 
\ba
\frac{m_{\rho}^2}{m_{\rho}^2+Q^2} &\rightarrow& 
\frac{m_{\rho}^2 + 8 \Gamma_{\rho} m_{\pi}/\pi} 
{m_{\rho}^2+Q^2 + (4m_{\pi}^2+Q^2) \Gamma_{\rho} \alpha(Q^2)/m_{\pi}} ~, 
\ea
with 
\ba
\alpha(Q^2) &=& \frac{2}{\pi} 
\left[ \frac{4m_{\pi}^2+Q^2}{Q^2} \right]^{1/2} 
\ln \left( \frac{\sqrt{4m_{\pi}^2+Q^2} + \sqrt{Q^2}}{2m_{\pi}} \right) ~.
\ea
The coefficients $\alpha^{S/V}$ and $\beta^{S/V}$ in Eq.~(\ref{SV}) 
are determined by the electric charges and the magnetic moments of the 
nucleon, respectively
\ba
\alpha^S &=& 1-\alpha_{\omega}-\alpha_{\phi} ~,
\nonumber\\
\alpha^V &=& 1-\alpha_{\rho} ~, 
\nonumber\\
\beta^S &=& 3(\mu_p + \mu_n) - \beta_{\omega}-\beta_{\phi} ~,
\nonumber\\
\beta^V &=& \frac{3}{5} (\mu_p - \mu_n) - \beta_{\rho} ~.
\ea
For small values of the momentum transfer the form factors are 
dominated by the meson dynamics and reduce to a monopole form, whereas 
for large values the modification of dimensional counting laws from 
perturbative QCD is taken into account by scaling 
$Q^2$ with the strong coupling constant \cite{GK,Speth} 
\ba
Q^2 &\rightarrow& Q^2 \frac{\alpha_s(0)}{\alpha_s(Q^2)} \;=\; 
Q^2 \frac{\ln [(Q^2 + \Lambda_{asy}^2)/\Lambda_{QCD}^2]} 
{\ln [\Lambda_{asy}^2/\Lambda_{QCD}^2]} ~. 
\ea

\subsection{Results}

\begin{figure}
\vfill 
\begin{minipage}{.48\linewidth}
\centerline{\psfig{file=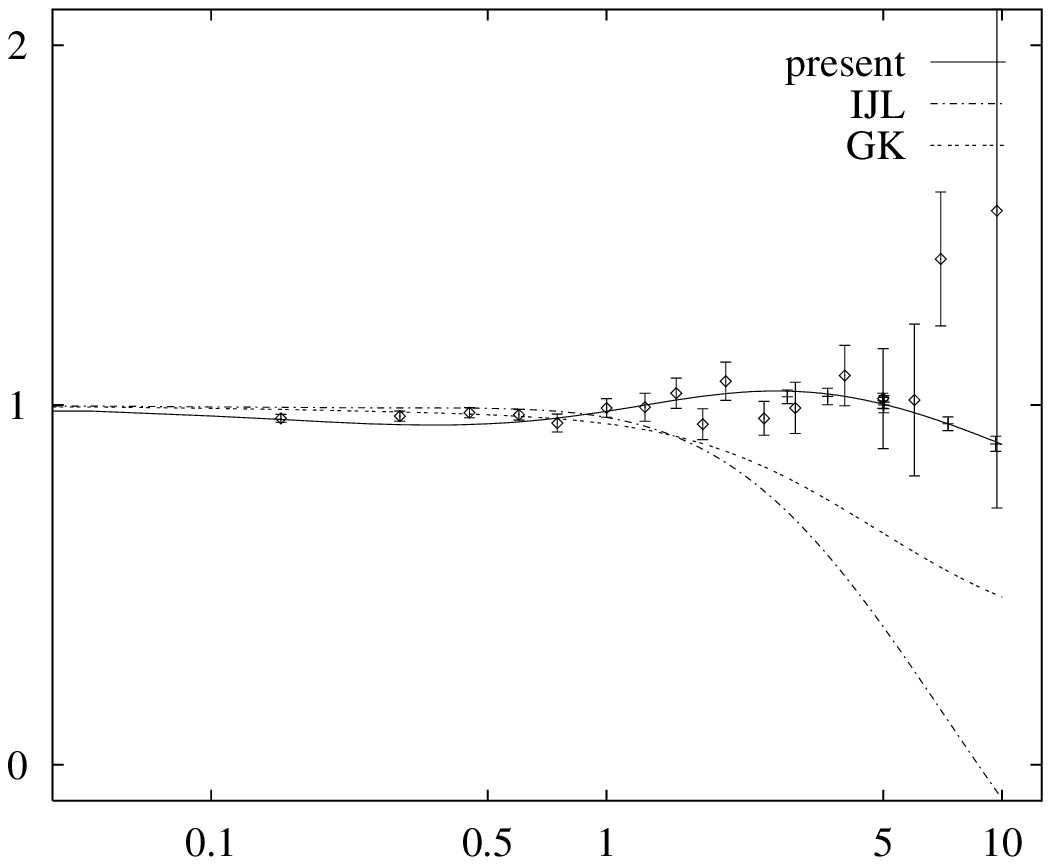,width=0.95\linewidth}}
\vspace{5pt}
\caption[]{Proton electric form factor 
$G_E^p/F_D$ as a function of $Q^2$ in (GeV/c)$^2$. 
The data are taken from \protect\cite{gpdat}.}
\label{gepfd}
\end{minipage}\hfill
\begin{minipage}{.48\linewidth}
\centerline{\psfig{file=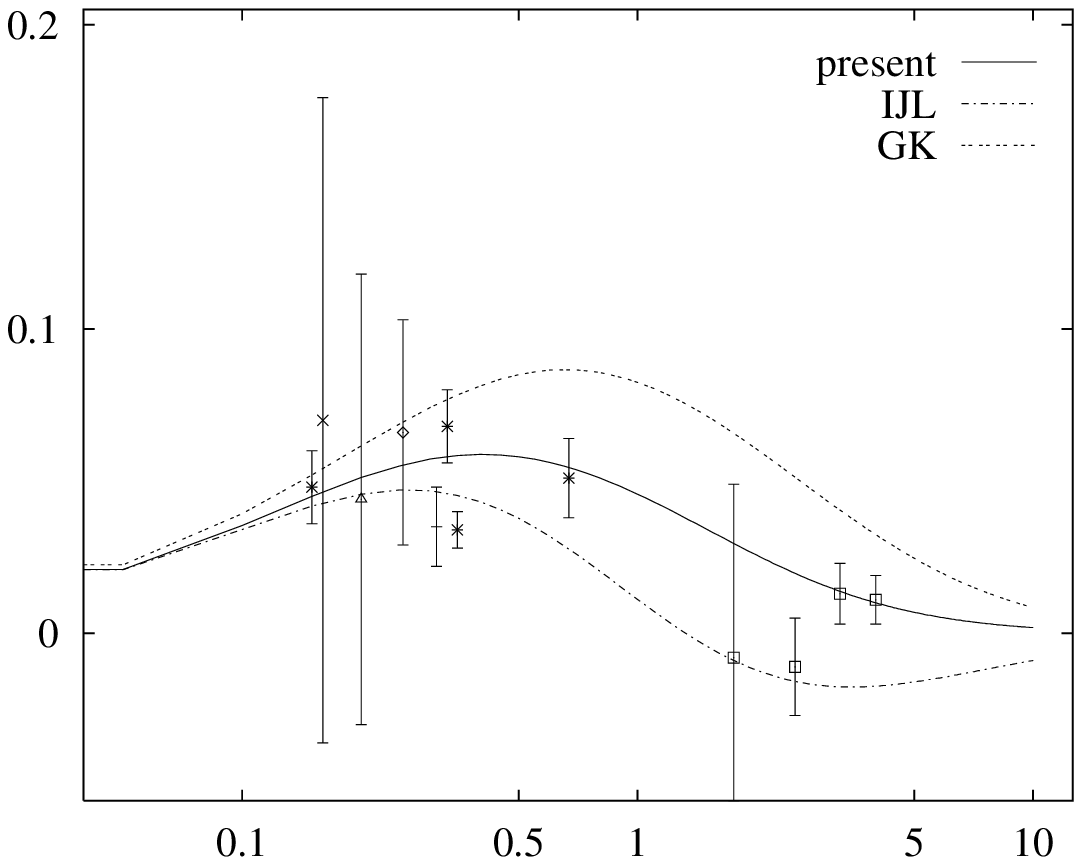,width=0.95\linewidth}}
\vspace{5pt}
\caption[]{Neutron electric form factor $G_E^n$ 
as a function of $Q^2$ in (GeV/c)$^2$. 
The data are taken from \protect\cite{gendat}.}
\label{gen}
\end{minipage}
\end{figure}

\begin{figure}
\vfill
\begin{minipage}{.48\linewidth}
\centerline{\psfig{file=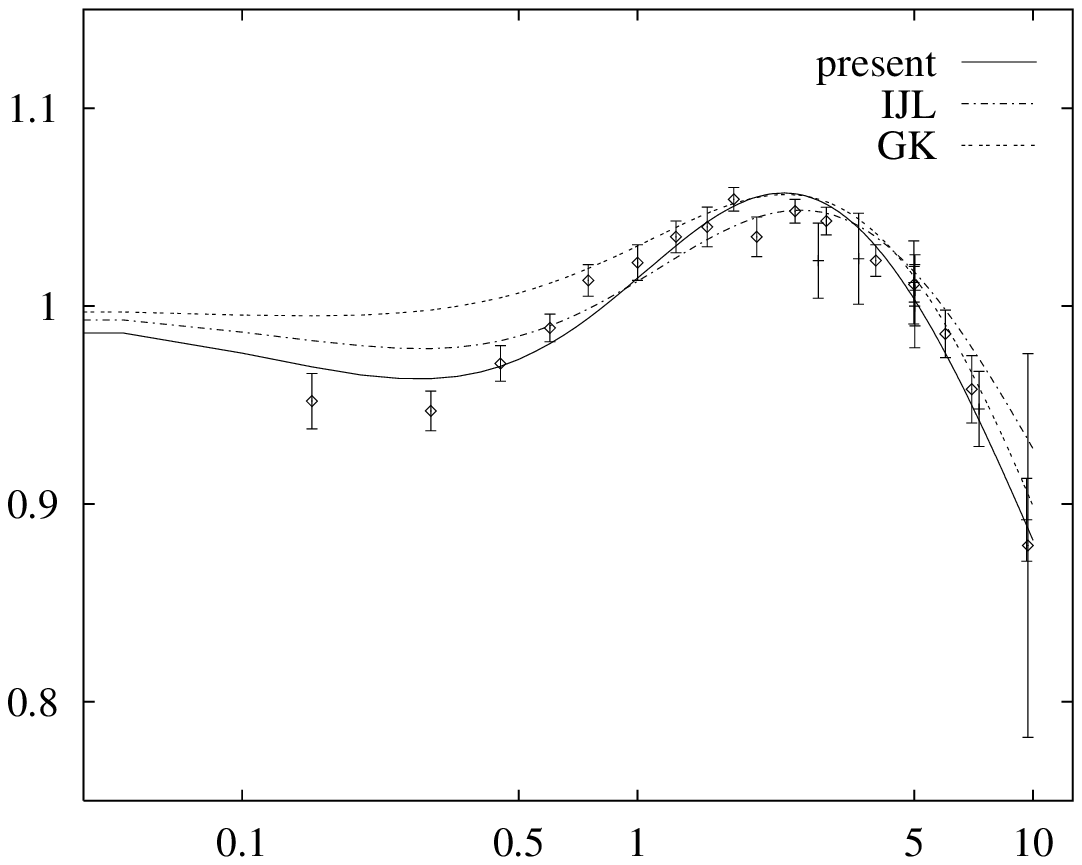,width=0.95\linewidth}}
\vspace{5pt}
\caption[]{Proton magnetic form factor $G_M^p/\mu_p F_D$ 
as a function of $Q^2$ in (GeV/c)$^2$. 
The data are taken from \protect\cite{gpdat}.}
\label{gmpfd}
\end{minipage}\hfill
\begin{minipage}{.48\linewidth}
\centerline{\psfig{file=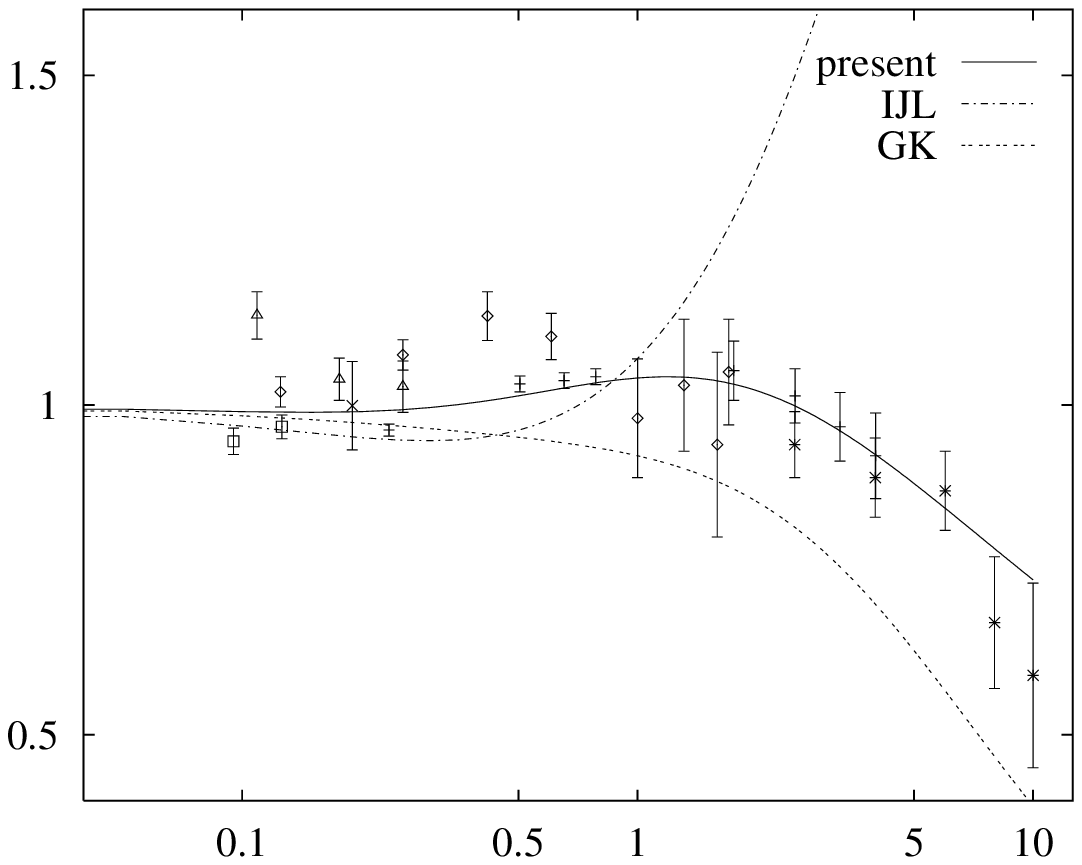,width=.95\linewidth}}
\vspace{5pt}
\caption[]{Neutron magnetic form factor $G_M^n/\mu_n F_D$ 
as a function of $Q^2$ in (GeV/c)$^2$. 
The data are taken from \protect\cite{gmndat1,gmndat2}.}
\label{gmnfd}
\end{minipage}
\end{figure}

In Figs.~\ref{gepfd}--\ref{gmnfd} we show a compilation of the most 
recent data on the electromagnetic form factors of the nucleon. 
Most of the data points have been obtained by a separation technique 
which is based on the Rosenbluth formula 
for the cross section \cite{Rosenbluth}
\ba
\frac{d \sigma}{d \Omega} &=&  
\left( \frac{d \sigma}{d \Omega} \right)_{Mott} \left[ 
\frac{G_E^2(Q^2) + \tau G_M^2(Q^2)}{1+\tau} 
+2\tau G_M^2(Q^2) \tan^2 \frac{\theta}{2} \right] ~, 
\label{sigma}
\ea
with $\tau=Q^2/4M^2$. Although in principle the electric 
and magnetic form factors can be separated by varying the 
scattering angle $\theta$, in practice this separation technique 
is limited to small values of $Q^2$ only. For large values 
of $Q^2$ the measured cross section is dominated by 
the magnetic form factor, and hence it becomes increasingly 
difficult to extract the electric form factor. For the neutron  
form factors there is the additional problem of the lack of a free 
neutron target, and the (small) relative size of the electric 
versus the magnetic form factors for small values of $Q^2$. 
In order to overcome these problems to extract the form factors 
many times the assumption of form factor scaling (which holds 
for small values of $Q^2$) is made 
\ba
G_E^p(Q^2) \;=\; \frac{G_M^p(Q^2)}{\mu_p} 
           \;=\; \frac{G_M^n(Q^2)}{\mu_n} ~. 
\label{scaling}
\ea
As a result, especially for the neutron form factors the scattering  
of data points is at times even larger than the uncertainties quoted 
by the authors. 
With the advent of high duty cycle electron facilities such as {\em e.g.} 
ELSA, MAMI and Jefferson Laboratory far more accurate determinations of 
the form factors become feasible, either using coincidence experiments 
(ratio method) or polarization variables \cite{gendat,gmndat2}. 
For example, the use of polarization variables makes it possible to 
extract the ratio of the electric and magnetic form factors from the data.

In the present calculation the coefficients $\alpha_M$ and $\beta_M$ 
(with $M=\rho$, $\omega$, $\phi$), the scale parameter $\gamma$ 
in the dipole form factor $g(Q^2)$ and the $\Lambda$'s 
are determined in a simultaneous fit to all four electromagnetic form 
factors of the nucleon and the proton and neutron charge radii. 
The values of the fitted parameters are given in Table~\ref{parameters}. 
As usual, the form factors are scaled by the standard dipole 
fit $F_D=1/(1+Q^2/0.71)^2$. The oscillations around the dipole 
values are attributed to the meson cloud couplings. 

\begin{figure}
\vfill 
\begin{minipage}{.48\linewidth}
\centerline{\psfig{file=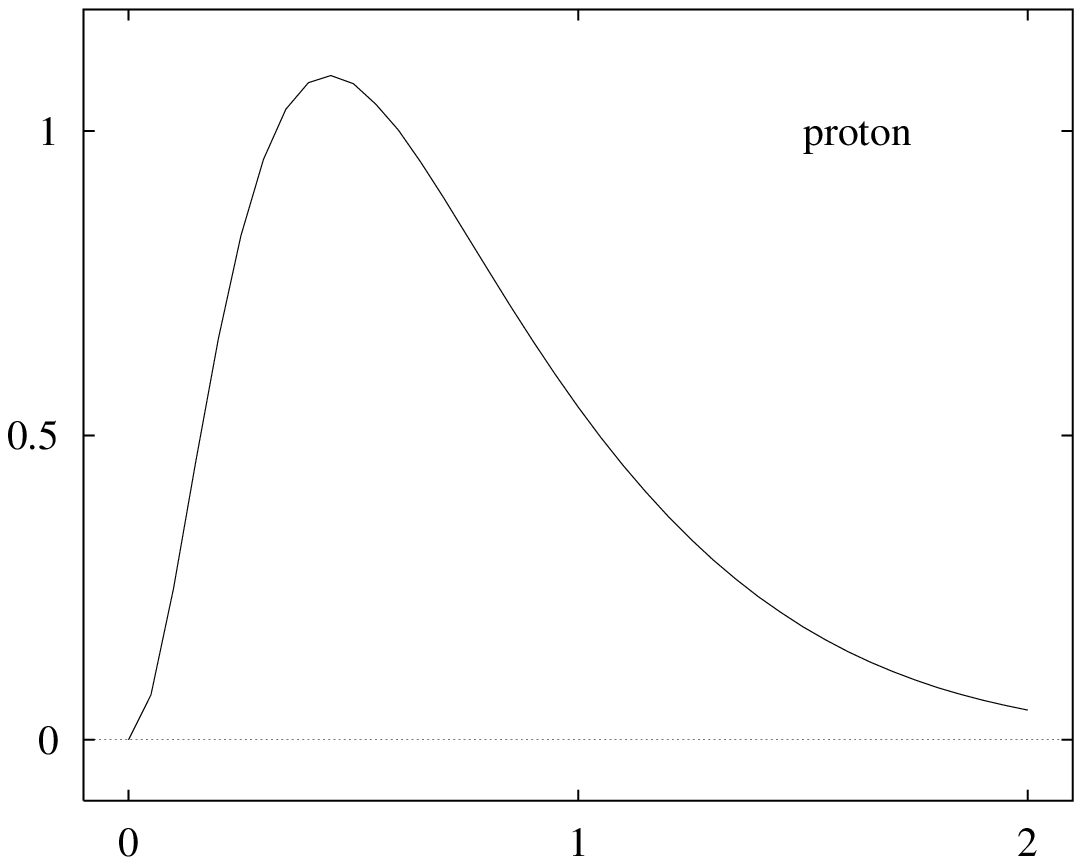,width=0.95\linewidth}}
\vspace{5pt}
\caption[]{Charge distribution of the proton 
$4 \pi r^2 \rho_p(r)$ as a function of r in fm.}
\label{rhop}
\end{minipage}\hfill
\begin{minipage}{.48\linewidth}
\centerline{\psfig{file=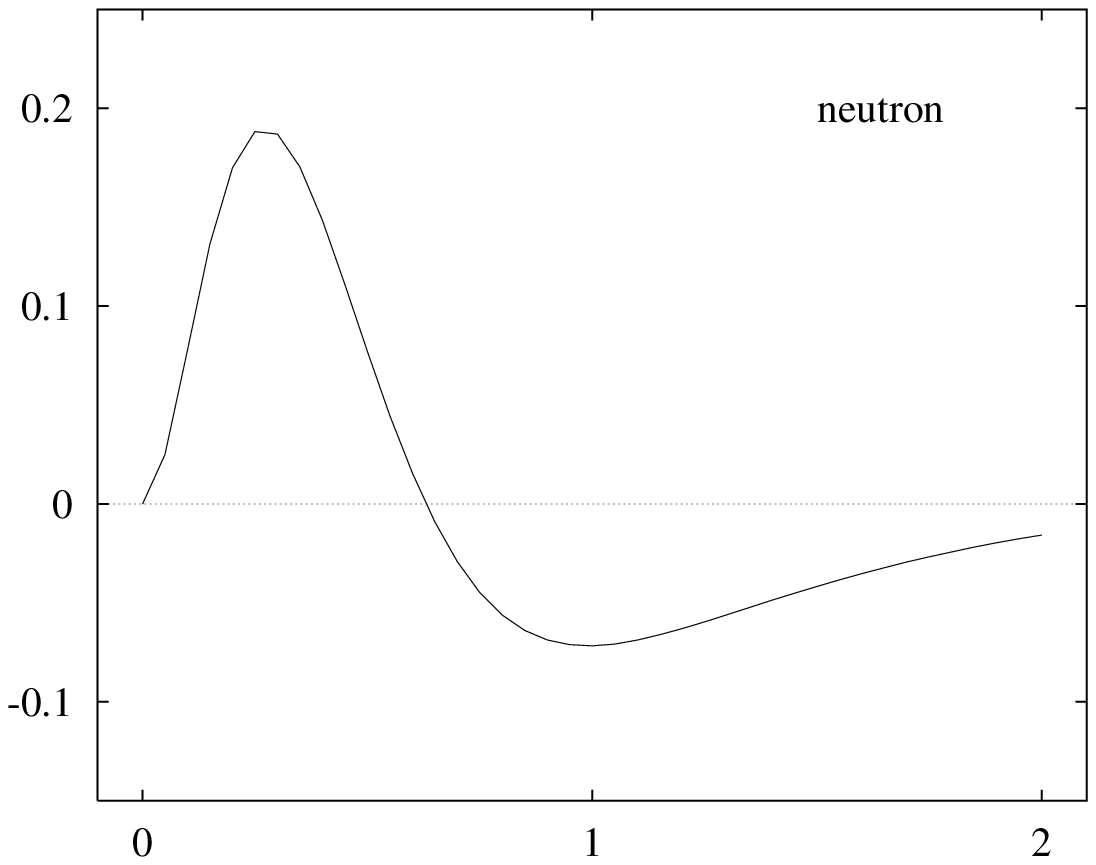,width=0.95\linewidth}}
\vspace{5pt}
\caption[]{Charge distribution of the neutron 
$4 \pi r^2 \rho_n(r)$ as a function of r in fm.}
\label{rhon}
\end{minipage}
\end{figure}

The electric form factors, as well as the proton and neutron charge 
radii (the slope of $G_E^p$ and $G_E^n$ in the origin) are reproduced 
well. The electric form factor of the neutron is the least known. 
Unlike for the proton, the Rosenbluth separation of $G_E^n$ from  
$G_M^n$ for a neutron target is difficult for all values of $Q^2$: 
for small $Q^2$ because of the small size of $G_E^n$ compared to 
$G_M^n$, and for large $Q^2$ because the magnetic component 
dominates both the angular dependent and angular independent 
term in the cross section. For this reason we have included 
in Fig.~\ref{gen} only the results obtained from polarization 
variables \cite{gendat}. The new data points for $G_E^n$ are 
significantly larger than those of the Platchkov compilation 
\cite{Platchkov}, which in turn leads to a reduction of the 
cross-over point in the neutron charge distribution from 0.9 fm 
to 0.7 fm \cite{Drechsel}. In the Breit frame, the proton and 
neutron charge distributions are given by the Fourier transforms 
of the respective electric form factors 
\ba
\rho_{p/n}(r) = \frac{1}{(2\pi)^3} \int d\vec{q} \, G_E^{p/n}(q) 
\, \mbox{e}^{-i \vec{q} \cdot \vec{r}} ~.
\ea
In Figs.~\ref{rhop} and \ref{rhon} we show the results of our 
calculations. The neutron charge distribution shows a change in 
sign at 0.65 fm.

The magnetic form factors show a slight oscillatory behavior 
around the dipole form. For large values of $Q^2$ both the proton  
and the neutron magnetic form factor show a decrease with respect to 
the dipole. The present calculation is in good agreement with the 
new measurements of $G_M^n$ at MAMI (Anklin '98 \cite{gmndat2}). 

\begin{figure}
\vfill 
\begin{minipage}{.48\linewidth}
\centerline{\psfig{file=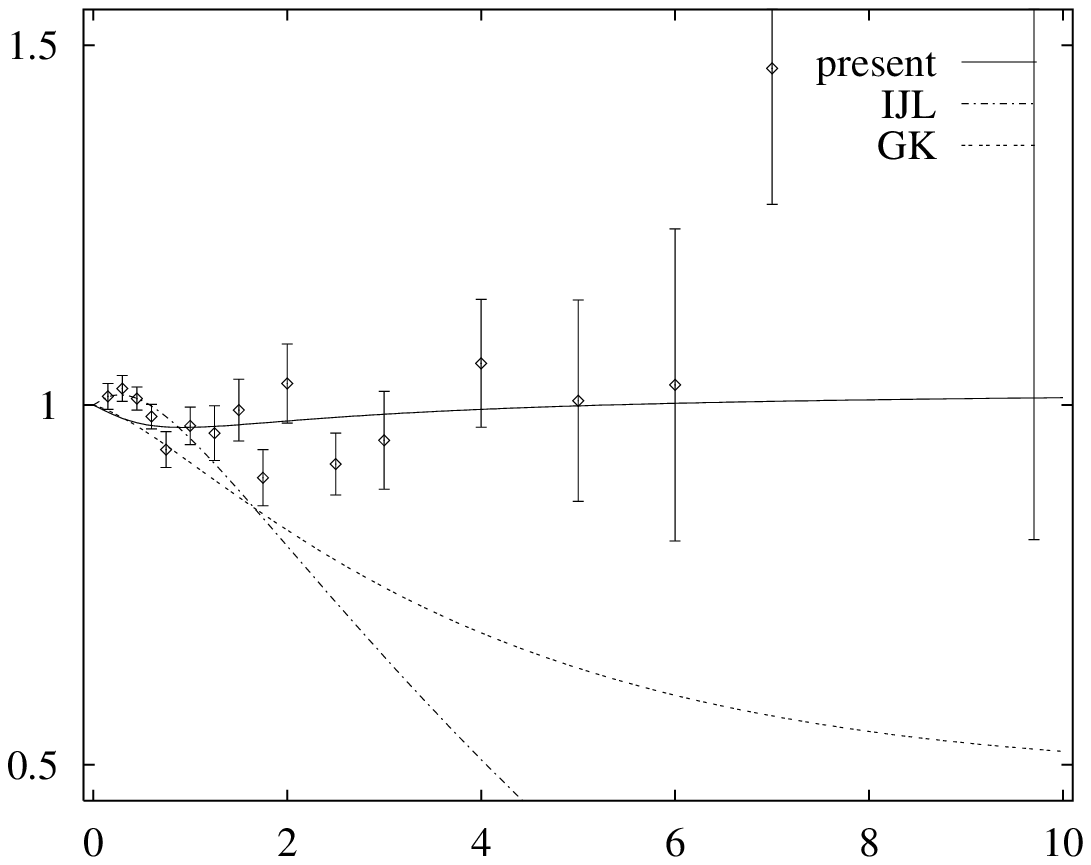,width=0.95\linewidth}}
\vspace{5pt}
\caption{Ratio of proton form factors $\mu_p G_E^p/G_M^p$ 
as a function of $Q^2$ in (GeV/c)$^2$. The data are taken 
from Walker et al. \protect\cite{gpdat}.}
\label{gepgmp}
\end{minipage}\hfill
\begin{minipage}{.48\linewidth}
\centerline{\psfig{file=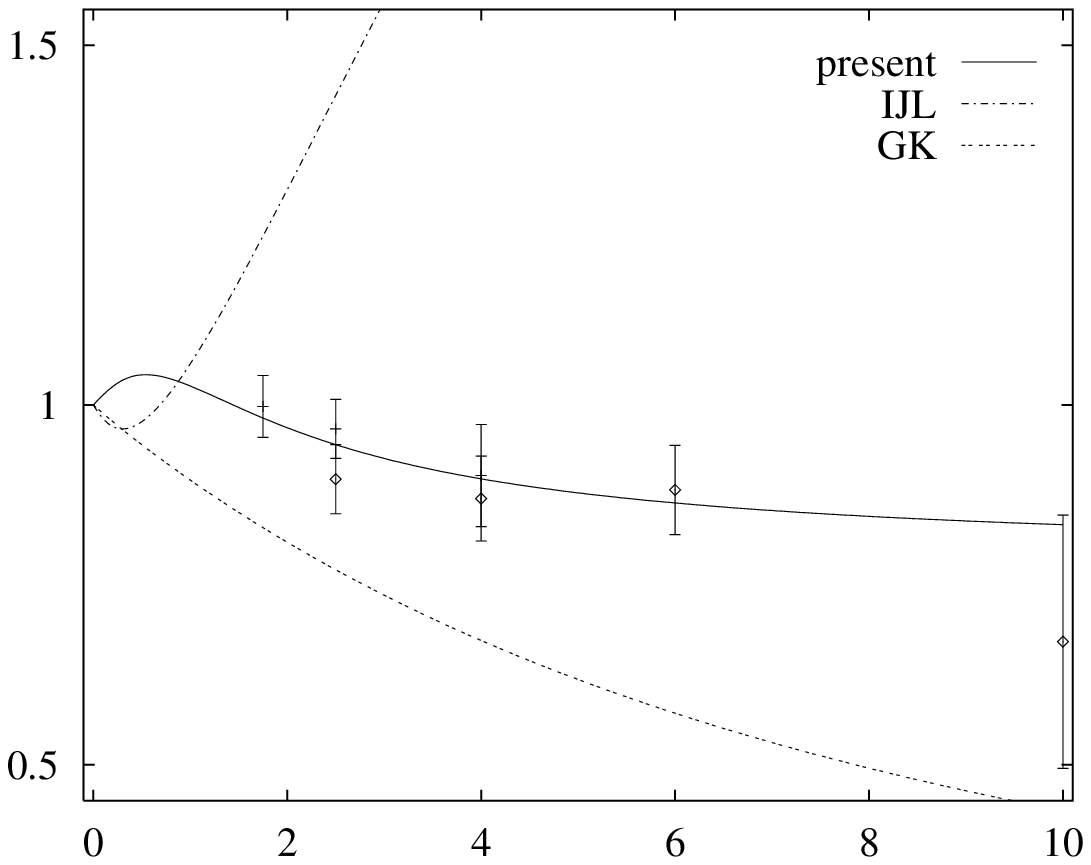,width=0.95\linewidth}}
\vspace{5pt}
\caption{Ratio of magnetic form factors $\mu_p G_M^n/\mu_n G_M^p$ 
as a function of $Q^2$ in (GeV/c)$^2$. The data are taken 
from Rock et al. and Lung et al. \protect\cite{gmndat1}.}
\label{gmngmp}
\end{minipage}
\end{figure}

\begin{figure}
\vfill 
\begin{minipage}{.48\linewidth}
\centerline{\psfig{file=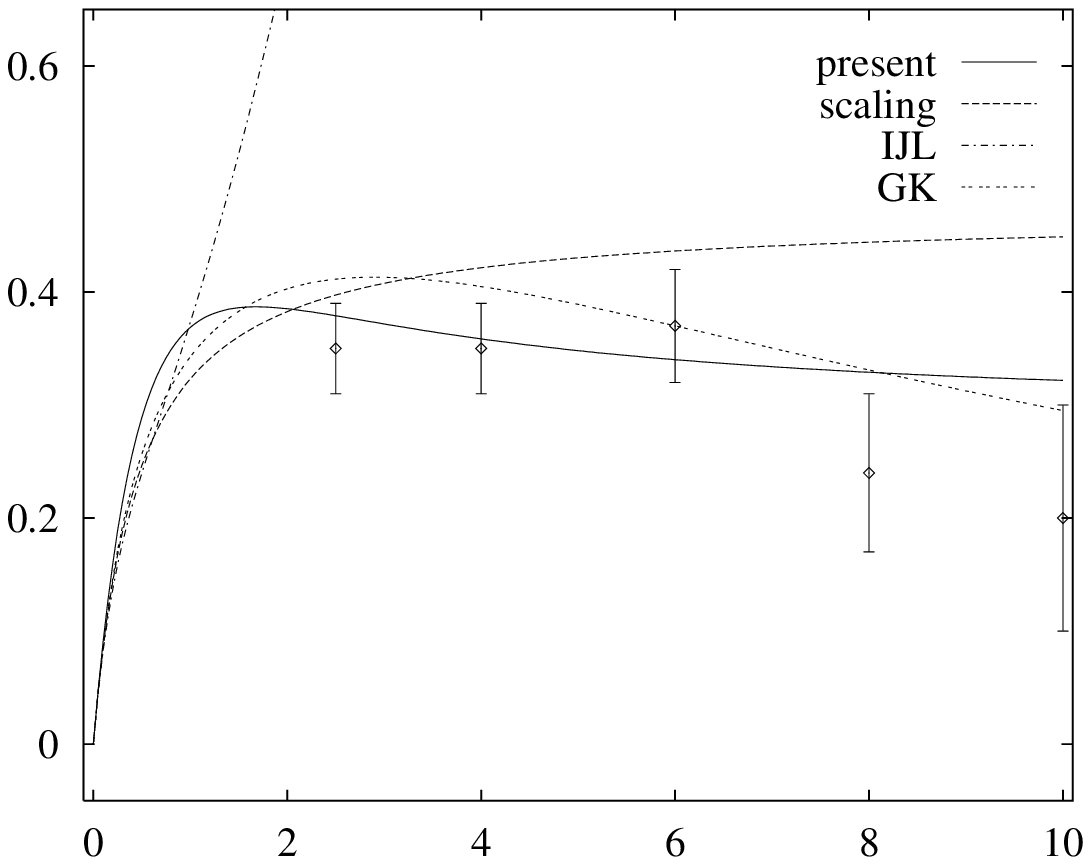,width=0.95\linewidth}}
\vspace{5pt}
\caption[]{Ratio of neutron and proton cross section of 
Eq.~(\ref{sigmanp}) as a function of $Q^2$ in (GeV/c)$^2$. 
The data are taken from Rock et al. \protect\cite{gmndat1}.}
\label{snsp}
\end{minipage}\hfill
\begin{minipage}{.48\linewidth}
\centerline{\psfig{file=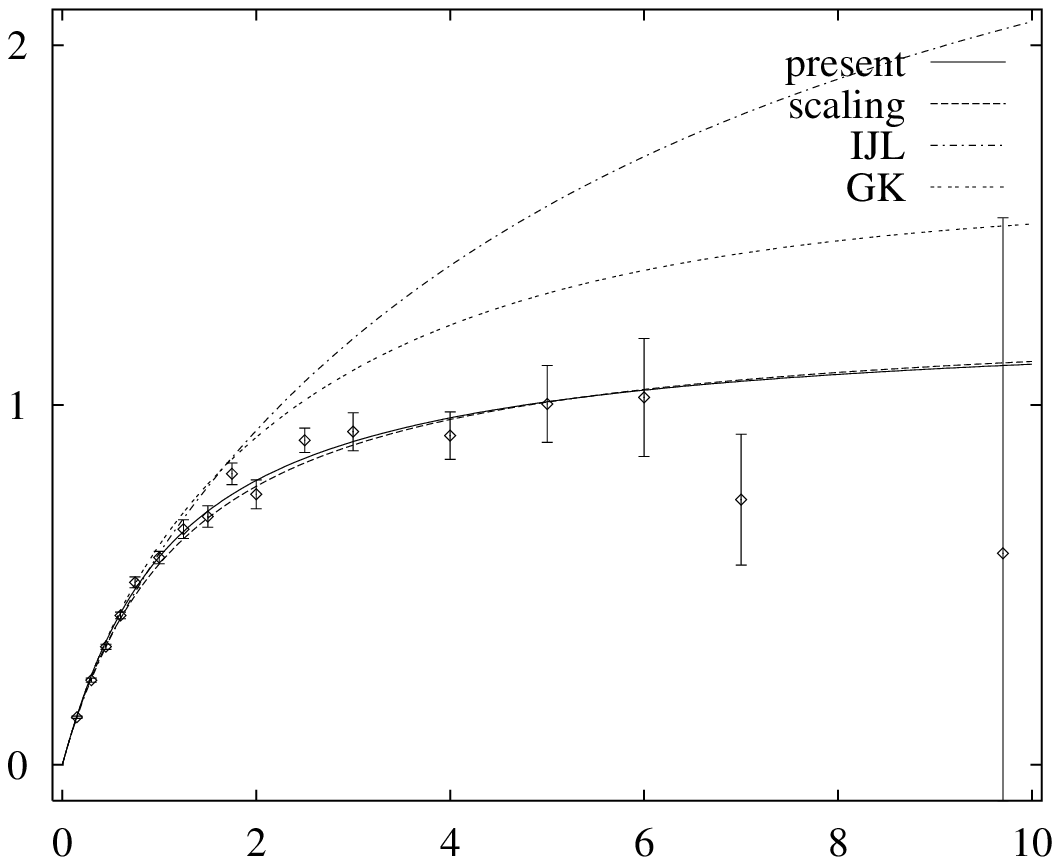,width=0.95\linewidth}}
\vspace{5pt}
\caption[]{Ratio of proton Pauli and Dirac form factors of 
Eq.~(\ref{f2f1}) in (GeV/c)$^2$ as a function of $Q^2$ in (GeV/c)$^2$. 
The data are taken from Walker et al. \protect\cite{gpdat}.}
\label{fprat}
\end{minipage}
\end{figure}

In Figs.~\ref{gepgmp} and \ref{gmngmp} we study the form factor scaling 
relations of Eq.~(\ref{scaling}). Whereas the ratio of the proton form 
factors is close to one over the entire range of $Q^2$, the ratio 
of neutron and proton magnetic form factors shows a deviation 
from form factor scaling.  
In Figs.~\ref{snsp} and \ref{fprat} we show two other measures 
of deviations from form factor scaling. The ratio of neutron and 
proton Rosenbluth cross sections reduces under the assumption of 
form factor scaling and for large values of $Q^2$ to a constant 
\ba
\left. \frac{d \sigma_n}{d \Omega} \right/
\frac{d \sigma_p}{d \Omega} 
&=& \frac{(G_E^n)^2 + \tau (G_M^n)^2 
+2\tau(1+\tau) (G_M^n)^2 \tan^2 (\theta_n/2) }
{(G_E^p)^2 + \tau (G_M^p)^2
+2\tau(1+\tau) (G_M^p)^2 \tan^2 (\theta_p/2)} 
\nonumber\\
&\rightarrow& \frac{\tau \mu_n^2 
+2\tau(1+\tau) \mu_n^2 \tan^2 (\theta_n/2)}
{1 + \tau \mu_p^2
+2\tau(1+\tau) \mu_p^2 \tan^2 (\theta_p/2)} 
\nonumber\\
&\rightarrow& \frac{\mu_n^2 \tan^2 (\theta_n/2)}
{\mu_p^2 \tan^2 (\theta_p/2)} ~.
\label{sigmanp}
\ea
For equal angles ($\theta_n = \theta_p$) this ratio reduces to 
$\mu_n^2/\mu_p^2 = 0.47$~. 
The transition from the region of low momentum transfer where 
to good approximation the nucleon form factors satisfy form factor 
scaling, to the region of high momentum transfer 
for which the methods of perturbative QCD apply, can be 
studied by the ratio $Q^2 F_2^p / F_1^p$ of the Dirac ($F_1$) and 
Pauli ($F_2$) form factors
\ba
F_1(Q^2) &=& \frac{G_E(Q^2) + \tau G_M(Q^2)}{1+\tau} ~,
\nonumber\\
F_2(Q^2) &=& \frac{G_M(Q^2) - G_E(Q^2)}{(\mu_p-1)(1+\tau)} ~.
\ea
According to dimensional scaling laws the helicity conserving 
amplitude $F_1^p$ dominates the helicity-flip amplitude $F_2^p$ 
at high $Q^2$, and the ratio $Q^2 F_2^p / F_1^p$ goes to a constant 
\cite{Brodsky}. Under the assumption of form factor scaling and for large 
values of $Q^2$ we find 
\ba
\frac{Q^2 F_2^{p}}{F_1^{p}} &=& 
\frac{Q^2(G_M^{p} - G_E^{p})}{(\mu_p-1)(G_E^{p} + \tau G_M^{p})} 
\nonumber\\
&\rightarrow& \frac{Q^2}{1 + \tau \mu_p} 
\nonumber\\ 
&\rightarrow& \frac{4M^2}{\mu_p} \;=\; 1.26 \mbox{ (GeV/c)}^2 ~.
\label{f2f1}
\ea
In Figs.~\ref{snsp} and \ref{fprat} both the data and our calculations 
show a saturation with increasing values of $Q^2$. A comparison 
between the full calculation (solid lines, labeled `present') and the 
assumption of form factor scaling (dashed lines, labeled `scaling') 
shows that for the ratio of the neutron and proton cross sections there 
is a large deviation from form factor scaling. For the ratio 
$Q^2 F_2^p/F_1^p$ the two curves coincide. 

For comparison we also show the results of two other calculations: 
the vector meson dominance model of Iachello, Jackson and Lande \cite{IJL}
(dash-dotted lines, labeled `IJL'), and a hybrid model (interpolation 
between vector meson dominance and pQCD) by Gari and Kr\"umpelmann 
\cite{GK} (dash-dashed lines, labeled `GK'). 
In comparing the different calculations 
one has to keep in mind that each one was optimized with the 
data set of that time (1973 for \cite{IJL} and 1985 for \cite{GK}). 

\section{Transition form factors}

In addition to the elastic form factors, there is currently much interest 
in the inelastic transition form factors. Recent experiments on 
eta-photoproduction $\gamma + N \rightarrow N^{\ast} + \eta$ have 
yielded valuable new information on the helicity amplitudes of the 
N(1520)$D_{13}$ and N(1535)$S_{11}$ resonances. 
In an Effective Lagrangian Approach these new experimental results 
were used to extract model independent ratios of photocouplings 
\cite{Nimai1,Nimai2}
\ba
\mbox{N(1535)}S_{11} &: \hspace{1cm}& 
\frac{A^n_{1/2}}{A^p_{1/2}} \;=\; -0.84 \pm 0.15 ~.
\nonumber\\
\mbox{}
\nonumber\\
\mbox{N(1520)}D_{13} &: \hspace{1cm}& 
\frac{A^p_{3/2}}{A^p_{1/2}} \;=\; -2.5 \pm 0.2 \pm 0.4 ~. 
\ea
These values are in excellent agreement with those of the collective 
algebraic model, $-0.81$ and $-2.53$, respectively \cite{BIL} 
(for the N(1535)$S_{11}$ resonance a mixing angle of $\theta = -38$ 
degrees was introduced).  
In Table~\ref{ratios} we compare these values with 
some model calculations. Whereas for the ratio $A^n_{1/2}/A^p_{1/2}$ of 
the N(1535)$S_{11}$ resonance there is little variation between the 
various theoretical results, for the ratio $A^p_{3/2}/A^p_{1/2}$ of 
the N(1520)$D_{13}$ resonance there is a large spread in values. 
Note also the large discrepancy between the photocouplings obtained 
from electro-photoproduction \cite{PDG} and the new values 
determined from eta-photoproduction \cite{Nimai1,Nimai2,Tiator}. 

\begin{figure}
\vfill 
\begin{minipage}{.48\linewidth}
\centerline{\psfig{file=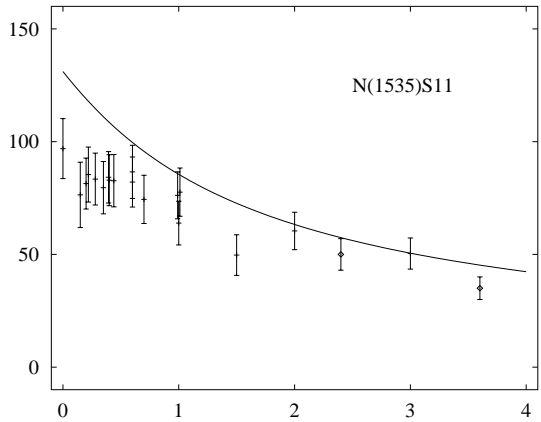,width=0.95\linewidth}}
\vspace{5pt}
\caption[]{N(1535)$S_{11}$ proton helicity amplitude in 
$10^{-3}$ GeV$^{-1/2}$ as a function of $Q^2$ in (GeV/c)$^2$. 
A factor of $+i$ is suppressed. The data are taken from 
a compilation in \protect\cite{Armstrong}.}
\label{n1535}
\end{minipage}
\end{figure}

Finally, in Fig.~\ref{n1535} we show the N(1535)$S_{11}$ proton helicity 
amplitude $A^p_{1/2}$ as a function of $Q^2$, for which there exist 
interesting new data \cite{Armstrong} (diamonds). The other points 
are obtained from a reanalysis of old(er) data, but now using the same 
values of the resonance parameters for all cases 
(from a compilation in \cite{Armstrong}). The solid curve represents 
the results of the collective algebraic model of \cite{BIL}, which 
were obtained by introducing a mixing angle of $\theta = -38$ degrees, 
but which do not contain the effects of meson-cloud couplings. 
We find good overall 
agreement with the data for the entire range of $Q^2$ values. 

\section{Summary and conclusions}

We presented a simultaneous analysis of the four elastic 
form factors of the nucleon and the transition form factors 
in a collective model of baryons, and found in general 
good agreement with the data. 

The elastic electromagnetic form factors of the nucleon were 
studied for the space-like region $0 \leq Q^2 \leq 10$ (GeV/c)$^2$. 
Whereas for low $Q^2$ the form factors satisfy to a good approximation 
form factor scaling, in the region of high $Q^2$ they exhibit 
deviations from these simple scaling relations, most notably for the 
ratio of the neutron and proton magnetic form factors and the 
neutron and proton cross sections. The deviations of the nucleon 
form factors at low $Q^2$ from the dipole 
form were attributed to couplings to the meson cloud. 

For a phenomenological 
approach (as the present one) a good data set is a prerequisite.  
Whereas the proton form factors are relatively well known, there is 
still quite some controversy about the neutron form factors. 
New measurements of the polarization asymmetry in which the ratio of 
the electric and magnetic form factor of the neutron is extracted 
may help to further clarify the experimental situation. 

In conclusion, the present analysis of electromagnetic 
couplings shows that the collective model of baryons 
provides a good overall description of the available data.  

\section*{Acknowledgements}

This work is supported in part by DGAPA-UNAM under project IN101997 
and by grant No. 94-00059 from the United States-Israel Binational 
Science Foundation (BSF), Jerusalem, Israel.

\clearpage

\begin{table}
\centering
\caption[]{Parameter values.}
\label{parameters}
\vspace{5pt}
\begin{tabular}{crl}
\hline
& & \\
Parameter & Fit & \\
& & \\
\hline
& & \\
$\gamma$          &   0.710 & GeV$^{-2}$ \\  
$\alpha_{\rho}$   &   0.371 \\
$\alpha_{\omega}$ &   0.787 \\
$\alpha_{\phi}$   & --0.677 \\
$\beta_{\rho}$    &   1.261 \\
$\beta_{\omega}$  &   1.171 \\
$\beta_{\phi}$    & --1.056 \\
& & \\
$\Lambda_{asy}$ & 1.439 & GeV \\
$\Lambda_{QCD}$ & 0.217 & GeV \\
& & \\
$\chi^2$          &  99/82 \\
& & \\
\hline
\end{tabular}
\end{table}

\begin{table}
\centering
\caption[]{Ratios of helicity amplitudes.}
\label{ratios} 
\vspace{5pt} 
\begin{tabular}{lcc}
\hline
& & \\
& N(1535)$S_{11}$ & N(1520)$D_{13}$ \\
& $A^n_{1/2}/A^p_{1/2}$ & $A^p_{3/2}/A^p_{1/2}$ \\
& & \\
\hline
& & \\
Feynman et al.   \cite{FKR}      & --0.69 & --3.21 \\
Koniuk and Isgur \cite{KI}       & --0.81 & --5.57 \\
Warns et al.     \cite{Warns2}   & --1.06 & --9.00 \\
Close and Li     \cite{CL}       & --0.74 & --6.55 \\
                                 & --0.65 & --4.87 \\
Li and Close     \cite{LC}       & --0.54 & --2.50 \\
                                 & --0.56 & --2.61 \\
Capstick         \cite{Capstick} & --0.83 & --8.93 \\
Bijker et al.    \cite{BIL}      & --0.81 & --2.53 \\
Santopinto et al. \cite{SIG}     & --0.68 & --1.55 \\
                                 & --0.67 & --1.80 \\
& & \\
Mukhopadhyay et al. \cite{Nimai1,Nimai2}  
& -0.84 $\pm$ 0.15 & -2.5 $\pm$ 0.2 $\pm$ 0.4 \\
PDG \cite{PDG}      & -0.51 $\pm$ 0.47 & -6.9 $\pm$ 2.8 \\
Tiator et al. \cite{Tiator} & & -2.1 $\pm$ 0.2 \\
& & \\
\hline
\end{tabular}
\end{table}

\clearpage

\end{document}